\documentclass[12pt]{article}
\usepackage{amsfonts,amssymb}
\newcommand{\p}[1]{(\ref{#1})}
\newcommand{\nn}{\nonumber}
\newcommand{\ba}{\begin{eqnarray}}
\newcommand{\ea}{\end{eqnarray}}
\newcommand{\be}{\begin{equation}}
\newcommand{\ee}{\end{equation}}
\newcommand{\bi}[1]{b_{#1}}
\newcommand{\bbi}[1]{{\bar b}_{#1}}
\newcommand{\ai}[1]{a_{#1}}
\newcommand{\abi}[1]{{\bar a}_{#1}}
\newcommand{\al}[1]{\alpha_{#1}}
\newcommand{\alb}[1]{{\bar\alpha}_{#1}}
\newcommand{\bt}[1]{\beta_{#1}}
\newcommand{\btb}[1]{{\bar \beta}_{#1}}

\begin{document}
\setcounter{page}{0}
 \thispagestyle{empty}

\begin{flushright}
{nlin.SI/0612064}\\
\end{flushright}

\vspace{2cm}

\begin{center}
{\LARGE {\bf Generalized N=4 supersymmetric   }}\\[0.3cm]
{\LARGE {\bf   Toda lattice hierarchy }}\\[0.3cm]
{\LARGE {\bf and N=4 superintegrable mapping}}\\[1cm]

\vspace{.5cm}

{\large V.V. Gribanov }

\vspace{.2cm}

{{\em Bogoliubov Laboratory of Theoretical Physics,}}\\
{{\em Joint Institute for Nuclear Research,}}\\
{\em 141980 Dubna, Moscow Region, Russia}~\quad\\
\end{center}

\vspace{1cm}

\centerline{{\bf Abstract}}

 \vspace{.2cm}

 \noindent It is  shown
that the one-dimensional generalized N=4 supersymmetric Toda
lattice (TL) hierarchy (nlin.Si/0311030) contains the N=4
super-KdV hierarchy with the first flow time   in the role of
space coordinate. Two different N=2 superfield forms of the
generalized N=4 supersymmetric TL equation, which are useful when
solving the N=4 super-KdV and (1,1)-GNLS hierarchies, are
discussed.

 \vspace{.8cm}


 \vspace{.2cm}


\vfill

{\em ~~~E-mail:} { gribanov@theor.jinr.ru}

\newpage

\section{Introduction}  The Toda lattice (TL) and its
supersymmetric extensions being one of the most important families
in the theory of integrable systems were the subject of many
studies for the last decades.    The remarkable property of the TL
equations is  that they are closely related to the famous
differential integrable hierarchies of the NLS and KdV types.
Recently,  the generalized N=4 supersymmetric TL hierarchy, which
contains the N=2 and N=4 supersymmetric TL equations as the result
of appropriate reductions,  was proposed \cite{my1}. Is is known
that the N=2   supersymmetric TL equation serves as the symmetry
transformation of the   N=2 supersymmetric NLS hierarchy
\cite{dls} while the N=4   supersymmetric TL equation is directly
relevant to the N=4 supersymmetric KdV hierarchy \cite{sor1,sor4}.
At the same time, the relevance of the generalized N=4
supersymmetric TL equation to the differential hierarchies has not
been studied yet and the present paper  addresses this problem.

One of the  methods for finding differential hierarchies starting
with the lattice equations  is based on solving the appropriate
symmetry equation \cite{lez1,lez2}. In this approach,
a number of know and new  N=2 supersymmetric  hierarchies of
differential equations were reproduced \cite{dls,ls1,ls2}.
Another procedure by which one can extract  differential hierarchy
from the lattice one associated with the  one-matrix model was
proposed in \cite{bonora} where it was demonstrated that the
lattice hierarchy contains already the differential one with the
first flow time $t_1$ in the  role of space coordinate. In such an
approach, all the flows of the lattice hierarchy can simply be
rewritten in the form of differential equations if one uses the
first flow of the lattice hierarchy in order to express all the
lattice fields via the lattice fields defined in the same lattice
node. In the present paper, we apply this approach to the
generalized N=4 supersymmetric TL hierarchy and demonstrate that
the generalized N=4 supersymmetric TL equation forms the discrete
symmetry of the N=4 supersymmetric KdV hierarchy.

The paper is organized as follows. In section 2, we recall the
Lax-pair representation of the generalized N=4 supersymmetric TL
hierarchy and consider two different reductions of its first flow
which lead to the N=2 and N=4 supersymmetric TL equations. In
section 3, we use the first flow of the the generalized N=4
supersymmetric TL hierarchy in order to rewrite all its flows in
terms of fields defined in the same lattice node, which allows us
to reproduce the component form of the N=4 supersymmetric KdV
hierarchy in some new basis as well as its supersymmetric
transformations and conservation laws. In section 4, we discuss
two different N=2 superfield representations of the generalized
N=4 supersymmetric TL equation which are useful when solving the
(1,1)-GNLS and N=4 supersymmetric KdV hierarchy.

\section{  1D generalized fermionic Toda lattice hierarchy}
In this section, we remind the Lax-pair formulation and the basic
properties of the one-dimensi\-onal generalized N=4 supersymmetric
TL hierarchy \cite{my1}. This hierarchy is generated by the
following equation:
 \ba \label{Lax}
\partial_k L=[(L_+)^k,L],
 \ea
 for the infinite supermatrices
 \ba\label{L-mat}
(L)_{i,j}=\delta_{i,j-2}+\gamma_i \delta_{i,j-1}+c_i
\delta_{i,j}+\rho_i \delta_{i,j+1} +d_i \delta_{i,j+2}, \ \ \ \
i\in\mathbb{Z} \ea
where $\partial_k\equiv\partial/\partial t_k$, the subscript $+$
denotes the upper (including diagonal) triangular part of the
matrix and the matrix entries $c_i, d_i$ ($\rho_i, \gamma_i$) are
the bosonic (fermionic) lattice fields with the Grassmann parity 0
(1) and the length dimensions $[d_i]=-2$, $[c_i]=-1$,
$[\rho_i]=-3/2$, $[\gamma_i]=-1/2$;   $i$  is the lattice index.
Note,  the supermatrix $L$ is bosonic and its Grassmann parity is
defined by the Grassmann parity of  elements $c_i$ on the main
diagonal.

The first two flows originated  from the Lax-pair representation
\p{Lax} have the following explicit form:
\ba \label{toda-1f} &&\partial_1 d_i=d_i(c_i-c_{i-2}),\ \ \ \ \
\partial_1 c_i=d_{i+2}-d_i+\gamma_i \rho_{i+1}+\gamma_{i-1}\rho_i, \nn  \\
&&\partial_1 \gamma_i=\rho_{i+2}-\rho_i,\ \ \ \ \
\partial_1 \rho_i=\rho_i (c_i-c_{i-1})+d_{i+1}
\gamma_i-d_i\gamma_{i-2}
\ea
and
 \ba \label{toda-2f}
\partial_2 d_i&=&d_i
(d_{i+2}-d_{i-2}+c_i^2-c_{i-2}^2-\rho_{i-2}\gamma_{i-3}
+\rho_{i-1}\gamma_{i-2}+\rho_{i}\gamma_{i-1}-\rho_{i+1}\gamma_{i}),\nn\\
\partial_2 c_i&=& d_{i+2} (c_i+c_{i+2}+\gamma_{i}\gamma_{i+1})-d_{i} (c_i+c_{i-2}+\gamma_{i-2}\gamma_{i-1})\nn\\
&&-\rho_i(\rho_{i-1}+\gamma_{i-1}(c_i+c_{i-1})) -\rho_{i+1}
(\rho_{i+2}+\gamma_{i}(c_i+c_{i+1})),\nn\\
\partial_2
\gamma_i&=&\rho_{i+2}(c_{i+1}+c_{i+2})-\rho_{i}(c_{i}+c_{i-1})+d_{i+3}\gamma_{i+2}
+(d_{i+2}-d_{i+1})\gamma_{i}-d_{i}\gamma_{i-2},\nn\\
\partial_2 \rho_i&=&\rho_i
(c_i^2-c_{i-1}^2+d_{i+2}-d_{i-1}-\rho_{i+1}\gamma_{i}-\rho_{i-1}\gamma_{i-2})\nn\\
&&-d_i (\rho_{i-2}+\gamma_{i-2}(c_{i-1}+c_{i-2}))+d_{i+1}
(\rho_{i+2}+\gamma_{i}(c_{i}+c_{i+1})).
 \ea

 Using the Lax pair representation (\ref{Lax}), it is easy
to derive the general expression for bosonic Hamiltonians, which
are in involution, via the standard formula
\begin{equation}\label{str}
H_k=\frac1k str L^k\equiv\frac1k \sum_{i=-\infty}^{\infty} (-1)^i
(L^k)_{ii}
\end{equation}
with the Hamiltonian densities $(-1)^i (L^k)_{ii}$  expected to
satisfy the equation with respect to the evolution time $t_s$
\ba\label{dl}
\partial_s ((-1)^i (L^k)_{ii})= \ell_{s,k,i}-\ell_{s,k,i-1}\equiv(\Delta\ell)_{s,k,i},\ \ \ \
\ea
where $\ell_{s,k,i}$  are polynomials of the lattice fields
$\{c_i,d_i,\gamma_i,\rho_i\}$. In what follows we assume the zero
boundary conditions at infinity for the lattice fields
\ba\label{bound} \lim_{i\rightarrow\pm
\infty}\{c_i,d_i,\gamma_i,\rho_i\}=0 \ea
in order the equation for the lattice conservation laws
\ba
\partial_s H_k= \sum_{i=-\infty}^{\infty}
(\Delta\ell)_{s,k,i}=\lim_{i\rightarrow\infty}\ell_{s,k,i}-\lim_{i\rightarrow
- \infty}\ell_{s,k,i}=0\ea
to be satisfied. Let us  give here the explicit expressions for
the first two lattice Hamiltonians obtained via formula \p{str}
 \ba\label{HAM-bos}
 &&H_1 = \sum_{i =- \infty }^{\infty} (-1)^i c_i,\ \ \ \ \ \ \ \
H_2 = \sum_{i =-\infty}^{\infty} (-1)^i (  \frac12 c_i^2 +d_i+
 \rho_i \gamma_{i- 1}),
 %
  \ea
The remarkable feature of the first flow \p{toda-1f} is that it
can be reduced to any of two 1D Toda lattice equations with
extended supersymmetry known up to recently. First, the N=4
supersymmetric TL equation can be deduced  by eliminating the
field $c_i$ from system \p{toda-1f} and   transition to the new
basis as follows:
\ba d_i=g_ig_{i-1},\ \ \ \ \rho_i=g_i\gamma_i^-,\ \ \ \
\gamma_i=\gamma_{i+1}^+. \ea
In this basis eq.\p{toda-1f} takes the following form
\ba \label{n4-toda}
\partial^2 \ln
g_i&=&g_{i+1}g_{i+2}-g_i(g_{i+1}-g_{i-1})+g_{i-1}g_{i-2}+g_{i+1}\gamma_{i+1}^+\gamma_{i+1}^--
g_{i-1}\gamma_{i-1}^+\gamma_{i-1}^-, \nn\\
\partial\gamma^\pm &=& g_{i+1}\gamma^\mp_{i+1}-g_{i-1}\gamma^\mp_{i-1},
\ea
that is a component form of the N=4 supersimmetric TL equation.

Second, the N=2 supersymetric TL equation   comes from  system
\p{toda-1f} as a result of the following reduction. Let us
introduce a new basis in the space $\{d_i,c_i,\rho_i,\gamma_i\}$
which separates odd and even lattice nodes
\ba \label{ab-basis}
 &&a_i=c_{2i+1},\ \ \ \ b_i=d_{2i+1},\ \ \ \
\alpha_i=\gamma_{2i-1},
\ \ \ \ \beta_i=\rho_{2i+1},\nn\\
&& \bar a_i= c_{2i},\ \ \ \ \bar b_i= d_{2i},\ \ \ \ \bar
\alpha_i=-\gamma_{2i}, \ \ \ \ \bar \beta_i=\rho_{2i}
\ea
and rewrite first flow \p{toda-1f} as  follows:
\begin{eqnarray}  \label{ab-toda}
&&\partial_1b_i=b_i(a_i-a_{i-1}),\ \ \
\partial_1a_i=
b_{i+1}-b_i+\beta_i\bar\alpha_i+ \alpha_{i+1}\bar\beta_{i+1},\nn \\
&&\partial_1\bar b_i=\bar b_i(\bar a_{i}-\bar a_{i-1}),\ \ \
\partial_1\bar a_i=
\bar b_{i+1}-\bar b_{i}+\beta_i\bar\alpha_i+ \alpha_{i}\bar\beta_{i},\nn\\
&&\partial_1\alpha_i=\beta_i-\beta_{i-1},\ \ \
 \partial_1\beta_i=(a_i-\bar a_i)\beta_i -b_i \alpha_i+\bar
b_{i+1}\alpha_{i+1},\nn\\
&&\partial_1\bar\alpha_i=\bar\beta_i-\bar\beta_{i+1},\ \ \
\partial_1\bar\beta_i=(\bar a_{i}- a_{i-1})\bar
\beta_i-b_i\bar\alpha_i+\bar b_{i}\bar\alpha_{i-1}.
 \end{eqnarray}
Now one can easily check that imposing  the reduction constraints
\ba\label{red_con}
 \bar b_i=0,\ \ \ \ \ \ \bar a_i=-\frac{\beta_i\bar\beta_i}{b_i}
\ea
on eq.\p{ab-toda} turns it into the following  system of
equations:
\begin{eqnarray}  \label{n2-toda}
&&\partial b_i=b_i(a_i-a_{i-1}),\ \ \
\partial a_i=
b_{i+1}-b_i+\beta_i\bar\alpha_i+ \alpha_{i+1}\bar\beta_{i+1},\nn \\
&& \partial\beta_i=a_i\beta_i -b_i \alpha_i,\ \ \
\partial \bar\beta_i=-a_{i-1}\bar \beta_i-b_i\bar\alpha_i, \nn \\
&&\partial \alpha_i=\beta_i-\beta_{i-1},\ \ \
\partial \bar\alpha_i=\bar\beta_i-\bar\beta_{i+1}.
\end{eqnarray}
which is recognized as a component form of the the N=2
supersymetric TL equation.

We would like to emphasize that eqs.\p{n4-toda} and \p{n2-toda}
are two different reductions of system \p{toda-1f} (or
\p{ab-toda}) which are not related to each other, i.e. N=2
supersymmetric TL equation \p{n2-toda} cannot be obtained from N=4
supersymmetric TL equation \p{n4-toda} as a result of some
reduction. Keeping in mind that both N=4 and N=2 TL equations can
be deduced from eq.\p{toda-1f} we call eq.\p{toda-1f} the
generalized N=4 supersymmetric TL equation.

The N=4 supersymmetry of  eq.\p{toda-1f} in the basis \p{ab-basis}
is realized by the following transformations:
\ba\label{toda-susy}
\delta_\epsilon b_i&=&\epsilon_1 (u_i\bar\beta_i+\bar
u_i\beta_i)-\epsilon_2 (u_i\bar\beta_i-\bar
u_i\beta_i)+\epsilon_3b_i(\alpha_i+\bar \alpha_i)+\epsilon_4
b_i(\alpha_i-\bar
\alpha_i),\nn\\
\delta_\epsilon\bar b_i&=&\epsilon_1(u_{i-1}\bar\beta_i+\bar
u_i\beta_{i-1})+\epsilon_2(u_{i-1}\bar\beta_i-\bar
u_i\beta_{i-1})+ \epsilon_3\bar b_i(\alpha_i+\bar \alpha_{i-1})+
\epsilon_4\bar b_i(\alpha_i-\bar \alpha_{i-1}),\nn\\
\delta_\epsilon a_i&=&\epsilon_1(\bar
u_{i+1}\alpha_{i+1}-u_{i}\bar\alpha_{i})- \epsilon_2(\bar
u_{i+1}\alpha_{i+1}-u_{i}\bar\alpha_{i})+
\epsilon_3(\beta_i-\bar\beta_{i+1})+\epsilon_4(\beta_i+\bar\beta_{i+1}),\nn\\
\delta_\epsilon\bar a_i&=&\epsilon_1(\bar
u_{i}\alpha_{i}-u_{i}\bar\alpha_{i})- \epsilon_2(\bar
u_{i}\alpha_{i}-u_{i}\bar\alpha_{i})+\epsilon_3(\beta_i-\bar\beta_{i})+
\epsilon_4(\beta_i+\bar\beta_{i}),\nn\\
\delta_\epsilon \beta_i&=& \epsilon_1  u_i (\bar
a_i-a_i)+\epsilon_2u_i (\bar a_i-a_i)
+\epsilon_3(b_i-\bar b_{i+1}-\beta_i\bar\alpha_i)+\epsilon_4(\bar b_{i+1}-b_i+\beta_i\bar\alpha_i),\nn\\
\delta_\epsilon \bar\beta_i&=&\epsilon_1\bar u_i (  a_{i-1}-\bar
a_i)- \epsilon_2\bar u_i (  a_{i-1}-\bar a_i)+
\epsilon_3(b_i-\bar b_i-\bar\beta_i\alpha_i) +\epsilon_4(b_i-\bar b_i-\bar\beta_i\alpha_i),\nn\\
\delta_\epsilon\alpha_i&=&\epsilon_1(u_{i-1}-u_{i})+\epsilon_2(u_{i-1}-u_{i})
+\epsilon_3(a_{i-1}-\bar a_i)+\epsilon_4(\bar a_i-a_{i-1}),\nn\\
\delta_\epsilon\bar\alpha_i&=&\epsilon_1(\bar u_{i+1}-\bar
u_{i})+\epsilon_2(\bar u_{i}-\bar u_{i+1}) +\epsilon_3(\bar
a_i-a_{i})+\epsilon_4(\bar a_i-a_{i}),
 \ea
where $\epsilon_k$ $(k=1,2,3,4)$ are the corresponding fermionic
infinitesimal parameters. Note that transformations corresponding
to the parameters $\epsilon_1$ and $\epsilon_2$ are nonlocal with
respect to the lattice indices and they are expressed via
composite fields   $u_i,\bar u_i$
 \ba\label{u-repr} u_i\equiv
\prod\limits_{k=0}^\infty
 {\displaystyle\frac{b_{i-k}}{\bar b_{i-k}}},
\ \ \ \ \ \ \ \bar u_i\equiv\prod\limits_{k=0}^\infty
 {\displaystyle\frac{\bar b_{i-k}}{ b_{i-k-1}}}
\ea
 which obey the following equations and N=4 supersymmetric
 transformations
\ba &&\partial_1 u_i=u_i(a_i-\bar a_i),\ \ \
\partial_1 \bar u_i=\bar u_i(\bar a_i- a_{i-1}),\nn\\
&&\delta_\epsilon
u_i=\epsilon_1\beta_i-\epsilon_2\beta_i+\epsilon_3u_i \bar
\alpha_i-\epsilon_4u_i\bar \alpha_i,\nn\\
&&\delta_\epsilon\bar u_i=\epsilon_1\bar\beta_i+\epsilon_2\bar
\beta_i+\epsilon_3\bar u_i \alpha_i+\epsilon_4\bar u_i \alpha_i.
\ea
It can be easily checked that the above transformations indeed
realize   N=4 supersymmetry, i.e., that their commutators for any
field $q_i\equiv\{a_i,\bar a_i,b_i,\bar
b_i,\beta_i,\bar\beta_i,\alpha_i,\bar\alpha_i \}$ give
\ba\label{com_susy}
 [\delta_\epsilon,\delta_{\tilde\epsilon}]q_i=2 \sum_{k=1}^4 (-1)^k {\tilde\epsilon}_k
 \epsilon_k \partial_1q_i.
 \ea

\section{From the Toda lattice  hierarchy to the differential one}
In this section, we show that   first flow \p{ab-toda} of the
generalized N=4 supersymmetric  TL hierarchy underlies the N=4
supersymmetric hierarchy of differtial equations which is related
to N=4 supersymmetric KdV hierarchy.

Our goal is to construct the hierarchy of differential equations
for which system \p{ab-toda} works as  the discrete symmetry
mapping connecting its different solutions. With this aim one can
deduce the appropriate symmetry equation \cite{lez1,lez2} and try
to solve it. However, in the case at hand we deal with all the
flows of the lattice hierarchy \p{Lax}, which allows us to apply
another approach   \cite{bonora} and simply rewrite all the
lattice flows via the lattice fields defined in the same lattice
node. Indeed, one can use   first flow \p{ab-toda} in order to
express all the lattice fields via the lattice fields defined in
the $i$-th lattice node as follows:
 \ba \label{subs}
 \begin{array}{ll}
 \ai{i-1}=\ai{i}-(\log\bi{i})',&
 \bt{i+1}=\bt{i}+\al{i+1}',\\
 \abi{i-1}=\abi{i}-(\log\bbi{i})',&
 \btb{i-1}=\btb{i}+\alb{i-1}',\\
 \bt{i-1}=\bt{i}-\al{i}',&
 \bi{i-1}=\bi{i}+\al{i}\btb{i}-\alb{i-1}\bt{i-1}-\ai{i-1}',\\
 \btb{i+1}=\btb{i}-\alb{i}',&
 \ai{i+1}=\ai{i}+(\log\bi{i+1})',\\
 \bbi{i+1}=\abi{i}'+\bbi{i}+\alb{i}\bt{i}-\al{i}\btb{i}, &
\abi{i+1}=\abi{i}+(\log\bbi{i+1})',\\
 \al{i+1}=(\bt{i}'+\bi{i}\al{i}-\bt{i}(\ai{i}-\abi{i}))/\bbi{i+1},~~~~~&
\alb{i+1}=(\bbi{i+1}\alb{i}+\btb{i+1}(\abi{i+1}-\ai{i})-\btb{i+1}')/\bi{i+1},\\
 \alb{i-1}=(\btb{i}'+\bi{i}\alb{i}+\btb{i}(\ai{i-1}-\abi{i}))/\bbi{i},&
\al{i-1}=(\bbi{i}\al{i}+\bt{i-1}(\ai{i-1}-\abi{i-1})-\bt{i-1}')/\bi{i-1},\\
 \bi{i+1}=\bi{i}-\al{i+1}\btb{i+1}+\alb{i}\bt{i}+\ai{i}',&
\bbi{i-1}=-\abi{i-1}'+\bbi{i}-\alb{i-1}\bt{i-1}+\al{i-1}\btb{i-1},
  \end{array}
 \ea
where we denote the  $t_1$-derivative by the sign $'$. With the
help of eq.\p{subs}
  all
the flows of the lattice hierarchy \p{Lax} being rewritten in
basis \p{ab-basis} can be expressed in the terms of the lattice
fields and their derivatives defined in the same lattice node.
%
%
Thus, one can verify that   second flow \p{toda-2f} turns into the
following system of differential equations:
\ba\label{ba-2f}
\partial_2 a&=&(a'+a^2+2 (b-\beta\bar\alpha))',\nn\\
\partial_2 \bar a&=&(\bar a'+\bar a^2+2 (\bar b+\bar\beta\alpha))',\nn\\
\partial_2 b&=&-b''+2 (a b)'+2 b (\bar\beta\alpha-\beta\bar\alpha),\nn\\
\partial_2 \bar b&=&-\bar b''+2 (\bar a \bar b)'+2  (\bar b \bar\beta\alpha+
( \alpha'-\beta)(\bar\beta'+b\bar\alpha+\bar\beta(a-\bar a-(\log b)'))),\nn\\
\partial_2 \beta&=&\beta''+2 (\bar a\,\beta+b\,\alpha)',\nn\\
\partial_2 \bar\beta&=&-\bar\beta''+2 (\bar a\,\bar\beta-b\,\bar\alpha)',\nn\\
\partial_2 \alpha&=&-\alpha''+2 (\beta'+\bar a\beta+\alpha(b-\bar b)
+( \alpha'-\beta)(a-(\log b)')),\nn\\
\partial_2 \bar\alpha&=&\bar\alpha''-2 (\bar \beta'-a\bar \alpha'+\bar\alpha(b-\bar b-\bar a')
+\bar\beta(a-\bar a-\alpha\bar\alpha) ),
 \ea
 where we omit the lattice index $i$.
Therefore, one can conclude that using substitutions \p{subs} one
can rewrite all the flows of N=4 TL hierarchy \p{Lax} in the form
of two-dimensional differential equations, i.e., pass from the
lattice hierarchy to the differential one. The composite lattice
fields \p{u-repr} which are nonlocal with respect to lattice
indices in this case become nonlocal  with respect to time
coordinate $t_1$ and take the following form:
 \ba
 u_i=exp(\partial^{-1}(a_i-\abi{i})):=e^\Delta, \ \ \ \ {\bar u}_i=\bi{i}
 exp(-\partial^{-1}(a_i-\abi{i})):=\bi{i} e^{-\Delta}
 \ea
The same procedure being applied to eq.\p{toda-susy} allows one to
find the explicit expressions for  the N=4 supersimmetric
transformations of the new differential hierarchy which read
\ba\label{new-susy}
 \delta_\epsilon a&=&\epsilon_1 ((b\,
\alpha-a\, \beta+\bar a\, \beta+\beta')e^{-\Delta}-\bar \alpha
e^{\Delta})+
\epsilon_2 ((a\, \beta-b\, \alpha-\bar
a\, \beta-\beta')e^{-\Delta}-\bar \alpha e^{\Delta} )\nn\\
&&+\epsilon_3 (\beta-\bar\beta+\bar\alpha')+\epsilon_4
(\beta+\bar\beta-\bar\alpha'),\nn\\
\delta_\epsilon \bar a&=&\epsilon_1 (b\, \alpha e^{-\Delta}-\bar
\alpha e^{\Delta})+
\epsilon_2 ( -b\, \alpha e^{-\Delta}-\bar \alpha e^{\Delta} )
+\epsilon_3 (\beta-\bar\beta )+\epsilon_4
(\beta+\bar\beta ),\nn\\
\delta_\epsilon b&=&\epsilon_1 (\bar \beta\,  e^{\Delta}+b\, \beta
e^{-\Delta})+
\epsilon_2 (\bar \beta\,  e^{\Delta}-b\,  \beta e^{-\Delta})
+\epsilon_3 b(\alpha+\bar\alpha )+\epsilon_4
b (\alpha-\bar\alpha ),\nn\\
\delta_\epsilon \bar b&=&\epsilon_1 (\bar b\, \bar \beta\,
e^{\Delta}/b+b\,( \beta-\alpha') e^{-\Delta})+
\epsilon_2 (\bar b\, \bar \beta\, e^{\Delta}/b-b\,(
\beta-\alpha') e^{-\Delta})\nn\\
&&+\epsilon_3 (\bar b\,\alpha+b\,\bar\alpha +\bar\beta (a-\bar
a-(\log b)' )+\bar\beta' )
+\epsilon_4 (\bar b\,\alpha-b\,\bar\alpha +\bar\beta (\bar a-
a+(\log b)' )-{\bar\beta}' ),\nn\\
\delta_\epsilon   \alpha&=&(\epsilon_1+\epsilon_2) (\bar b/b-1)
e^{\Delta}+(\epsilon_3-\epsilon_4)(a-\bar a-(\log b)'),\nn\\
\delta_\epsilon  \bar \alpha&=&(\epsilon_1-\epsilon_2) (\bar
b-b-\beta\bar\alpha+\bar\beta\alpha+{\bar a}')
e^{-\Delta}+(\epsilon_3+\epsilon_4)(\bar a-a),\nn\\
\delta_\epsilon  \beta&=&(\epsilon_1+\epsilon_2) (\bar a-a)
e^{\Delta}+(\epsilon_3-\epsilon_4)(b-\bar b-\bar\beta\alpha-\bar a'),\nn\\
\delta_\epsilon \bar \beta&=&(\epsilon_1-\epsilon_2) (b( a-\bar
a)-b') e^{-\Delta}+(\epsilon_3+\epsilon_4)(b-\bar
b-\bar\beta\alpha ).
 \ea
Note that the nonlocality of the supersymmetric transformations
corresponding to the parameters $\epsilon_1$ and $\epsilon_2$
originates from the nonlocality of the composite lattice fields
\p{u-repr}.

The main property of   integrable hierarchy is that it possesses
an infinite number of  conservation laws. From eq.\p{ba-2f} one
can find the first two conservation laws which have the following
simple form:
\ba\mathcal{H}_1=\int dx a, \ \ \ \ \ \ \ \ \ \ \ \ \
\bar\mathcal{H}_1=\int dx \bar a. \ea
Now let us show how all bosonic  conservation laws of the new
  N=4 supersymmetric differential hierarchy can be
produced from the lattice Hamiltonians \p{str}. By construction
the densities $(-1)^i (L^k)_{ii}$ are conserved which means
\ba\label{cc}
\partial_1\int dx (-1)^i
(L^k)_{ii}=\int dx (\ell_{1,k,i}-\ell_{1,k,i-1})=0, \ea
where we rename $t_1$ by $x$  and take into consideration
 the boundary conditions \p{bound} and eq.\p{dl}. From \p{cc} one
can obtain the relation   which connects the integrals
  defined in the neighboring lattice nodes and using
which   one can write
\ba \int dx \ell_{1,k,i}=\int dx \ell_{1,k,i-1}=\int
dx\lim_{i\rightarrow-\infty} \ell_{1,k,i}=0,
 \ea
where the boundary conditions \p{bound} are  taken into
consideration again. Thus, one can conclude that $\ell_{1,k,i}$ is
the conserved density and
\ba\label{laws} \mathcal{H}_{k+1}=\int dx \ell_{1,k,i} \ea
are the conservation laws, $\partial_s\mathcal{H}_k=0$. In such a
way, using eq.\p{subs} in order to express all the fields entering
into the density $\ell_{1,k,i}$ in the same lattice node we obtain
from Hamiltonians \p{HAM-bos} the next two conservation laws of
the differential N=4 supersymmetric hierarchy which read
\ba \mathcal{H}_2=\int dx (b-\bar b-\bar\beta\alpha), \ \ \ \ \ \
\ \
 \mathcal{H}_3=\int dx (ba-\bar b\bar a-b\alpha\bar\alpha-\bar
 a\bar\beta\alpha+\beta\bar\beta).
 \ea

 We have no   independent formulation of the new differential
 hierarchy so far. All its flows, the supersymmetry transformations as
 well as the conservation laws can be generated only with the help of the substitutions \p{subs}
 starting with the corresponding lattice counterparts.
In order to give an independent formulation of the new hierarchy,
one can try to find its Hamiltonian structure or construct its
Lax-pair representation. But first of all, to understand what kind
of hierarchy we deal with, let us consider the bosonic limit of
its second flow \p{ba-2f}. It is the system of two decoupled NLS
equations
\ba\label{ab-bos}
\partial_2 a&=&(a'+a^2+2 b)',\nn\\
\partial_2 b&=&(-b'+2 a b)'
 \ea
and the same equations for the fields $\bar a $ and $ \bar b$. The
set of equations \p{ab-bos} forms the bosonic limit of the second
flow of the a=4, N=2 supersymmetric KdV hierarchy. Keeping in mind
that the new hierarchy possesses N=4 supersymmetry one can expect
that it is closely related to the N=4 KdV hierarchy.
It turns out that such a relation indeed exists.
 After passing to the new basis
$\{u,v,r,s,\xi,\bar\xi,\eta,\bar\eta\}$ in the space of fields
$\{b,\bar b,a,\bar a,\alpha,\bar\alpha,\beta,\bar\beta\}$ defined
by the following transformations:
\ba \label{tr-a4}
&&u=b-\bar b-a'+(\log b)''+\alpha\bar\beta,\ \ \
\ v=a-(\log b)',\ \ \ \
 r=b\ e^{-\Delta},\nn\\
 &&s=-\bar b\ e^{\Delta}/b,\ \ \ \ \bar\xi=-\bar \beta,\ \ \ \
 \xi=\beta-\alpha',\ \ \ \ \eta=\alpha b\
 e^{-\Delta},\nn\\
 &&\bar\eta=((\bar a-a+(\log b)')\bar \beta -\bar \beta' -b\bar \alpha)\ e^{\Delta}/b
 \ea
one can rewrite the second flow \p{ba-2f} as follows:
 \ba \label{a4-kdv}
&&\partial_2 u= (-u'+ 2u v+2 r's-2 \xi\bar\xi+2 \eta\bar\eta)',\ \ \ \ \partial_2 \xi=(-\xi'-2 s\eta+2v\xi)',\nn\\
&&\partial_2 v= (v'+v^2+2 u-2 rs)',\ \ \ \ \partial_2 {\bar\xi}=(\bar\xi'-2 r\bar\eta+2v\bar\xi)',\nn\\
&&\partial_2 r= r''-2ur+2vr'+2 \eta\bar\xi,\ \ \ \
\partial_2\eta=
(\eta'+2r\xi+2v\eta)',\nn\\
&&\partial_2 s=-s''+2us+2 (vs)'-2  \xi\bar\eta,\ \ \ \
 \partial_2{\bar\eta}= (-\bar\eta'+2s\bar\xi+2v\bar\eta)'.
 \ea
The set of equations \p{a4-kdv} represents the component form of
the second flow of the a=4, N=4 supersymmetric KdV hierarchy
\cite{di,dik}. In \cite{sor4}, it was demonstrated that the a=4,
N=4 super-KdV hierarchy as well as the a=-2, N=4 super-KdV one can
be reproduced as a result of different reductions of the  N=4
Toda-KdV hierarchy written in  terms  of two constrained N=4
superfields. The component form of the second flow of the a=-2,
N=4 KdV hierarchy is \footnote{ Equation \p{a2-kdv} corresponds to
the second flow of   the a=-2, b=-6, N=4 super-KdV hierarchy,
according to the notation of \cite{dik}; the system \p{a4-kdv}
corresponds to the second flow of the a=4, b=0, N=4  super-KdV
hierarchy.}
\ba \label{a2-kdv}
\partial_2\tilde u&=&(\tilde r''+
\tilde u(\tilde r+3\tilde s)-\tilde v \tilde r'-3
\tilde\xi\tilde{\bar\eta}+\tilde{\bar\xi}\tilde\eta)',\nn\\
\partial_2\tilde v&=&\tilde s''-\tilde r''+2 \tilde u(\tilde r-\tilde s)+
\tilde v (\tilde r+\tilde s)'+ \tilde v'(\tilde s+3 \tilde r)+2
(\tilde\xi\tilde{\bar\eta}+\tilde{\bar\xi}\tilde\eta),\nn\\
\partial_2\tilde s&=&  \tilde u'+\tilde v''-\tilde u\tilde v- \tilde v\tilde v'+
\tilde s'(\tilde r+3\tilde s)-
\tilde\eta\tilde{\bar\eta}+\tilde\xi\tilde{\bar\xi},\nn\\
\partial_2\tilde r&=& \tilde u'+\tilde u\tilde v+
\tilde r'(\tilde s+3\tilde r)+
\tilde\eta\tilde{\bar\eta}-\tilde\xi\tilde{\bar\xi},\nn\\
\partial_2\tilde \xi&=& (-\tilde\eta'+\tilde v\tilde\eta+\tilde\xi(\tilde r+3 \tilde
s))',\ \ \ \ \
\partial_2\tilde {\bar\xi}\ =\ (\tilde{\bar\eta'}+\tilde { v}\tilde{\bar\eta}+
\tilde{\bar\xi}(\tilde { s}+3 \tilde { r}))',\nn\\
\partial_2\tilde \eta&=& (-\tilde\xi'-\tilde v\tilde\xi+\tilde\eta(\tilde s+3 \tilde
r))',\ \ \ \ \
\partial_2\tilde {\bar\eta}\ =\ (\tilde{\bar\xi'}-\tilde { v}\tilde{\bar\xi}+
\tilde{\bar\eta}(\tilde { r}+3 \tilde { s}))',
 \ea

One can easily verify that the component forms of the N=4
supersymmetric a=-2 and a=4 KdV hierarchies are related to each
other by the following simple
  relation:
\ba && \tilde u=u+s'+v'/2+r'/4,\ \ \ \tilde v=-2 s- r/2,\ \ \
\tilde
r=v/2-s+r/4,\ \ \ \tilde s=v/2+s-r/4,\nn\\
&&\tilde \xi=\xi-\eta/2,\ \ \ \tilde{\bar\xi}=\bar\xi/2- \bar
\eta,\ \ \ \tilde \eta=\eta/2+\xi,\ \ \ \tilde{\bar\eta}=\bar\eta
+\bar\xi/2.\ea

 Thus, one can conclude that the differential hierarchy deduced
 from the generalized N=4 supersymmetric TL hierarchy \p{Lax}
 with the help of
 substitutions \p{subs} is the component form of the N=4 super-KdV
 hierarchy.
  To finish this section,
  we   note that transformations \p{tr-a4} are invertible
\ba \label{tr-a4-inv} &&a=v+(\log(u+v'-rs-\bar\xi\eta/r))',\ \ \ \
b=u+v'-rs-\bar\xi\eta/r, \nn\\
&& \bar a=v+(\log r)',\ \ \ \  \bar b=-rs,\ \ \ \ \beta=\xi+(
\eta/r)',\ \ \ \ \bar \beta=-\bar\xi,
\nn\\
&&\alpha= \eta/r,\ \ \ \ \bar \alpha=\frac{\bar\xi'-\bar\xi(\log
r)'-r\bar\eta}{u+v'-rs-\bar\xi\eta/r}, \ea
and relations \p{tr-a4-inv} being substituted into eq.\p{ab-toda}
  give the discrete
symmetry mapping connecting different solutions of the a=4, N=4
super-KdV hierarchy.

\section{From the differential hierarchy to the lattice one}
In the previous  section, we demonstrated how differential N=4
supersymmetric KdV hierarchy  can be recovered from the
generalized N=4 supersymmetric  TL hierarchy \p{Lax}. In this
section, we resolve the inverse problem, i.e., show how  TL
hierarchy \p{Lax} can be reproduced from the differential one.

First of all, let us introduce the (1,1)-Generalized Nonlinear
Schr$\ddot{o}$dinger ((1,1)-GNLS) hierarchy \cite{bks}. All its
flows can be deduced from the Lax-pair representation
\ba\label{gnls}
\partial_k L=[(L^k)_{\geq 1},L]
\ea
for the Lax operator
\ba L=\partial-1/2 (F_a \overline F_a+F_a\overline D
\partial^{-1}[D \overline F_a]), \ea
where $F_a(X)$ and $\overline F_a(X)$ are chiral and antichiral
N=2 superfields
\ba DF_a(X) =0,\ \ \ \ \overline D\ \overline F_a(X) =0, \ea
 respectively, which are bosonic for $a=1$ and fermionic for
 $a=2$; $X=(x,\theta,\bar\theta)$ is a coordinate of N=2
 superspace and $D,\overline D$ are the N=2 supersymmetric fermionic
 covariant derivatives
 \ba
 D=\frac{\partial}{\partial\theta}-\frac12 \bar\theta \frac{\partial}{\partial
 x},\ \ \ \
\overline D=\frac{\partial}{\partial\bar\theta}-\frac12  \theta
\frac{\partial}{\partial
 x},\ \ \
\
 \{D,\overline D\}=-\frac{\partial}{\partial
 x}, \ \ \ D^2=\overline D^2=0
 \ea
 In \p{gnls} the subscript ${\geq 1}$ means the sum of purely
 derivative terms of the operator $L^k (k>0)$
 and for $k=2$ eq.\p{gnls} gives
\ba \label{gnls-2f}
 \partial_2 F_a&=& F_a{}''+D( F_b \overline F_b \overline DF_a),\nn \\
\partial_2 {\overline F}_a&=&-\overline F_a{}''+\overline D( F_b \overline F_b  D\bar
F_a),
 \ea
where summation over repeated indices is understood. The set of
equations \p{gnls-2f} forms the   (1,1)-GNLS equation which is
related to the a=4, N=4 supersymmetric KdV hierarchy. In
\cite{sorin}, it was demonstrated that the passing to the new
basis
\ba \label{1b}
 &&\mathcal{J}=-\frac{1}{2}\sum_{k=1}^2 F_k\overline F_k -\frac{
D\overline F_2' }{D\overline F_2},\ \ \ \ \
\mathcal{F}= - \frac12 F_1 D\overline F_2,\ \ \ \ \
 \overline{\mathcal{F}}=-\overline{D}D\left(
\frac{\overline F_1}{D\overline F_2}\right) \ea
establishes the relationship between these hierarchies. It is easy
to show that the new superfields $\mathcal{J}, \mathcal{F},
\overline\mathcal{F}$ satisfy, as a consequence of eq.\p{gnls-2f},
the following set of equations:
\ba \label{n4-kdv}
&&\partial_2 \mathcal{J}=([D,\overline D]\mathcal{J}+\mathcal{J}^2-2 \mathcal{F} \overline{\mathcal{F}})',\nn\\
&&\partial_2 \mathcal{F}=\mathcal{F}''-2 D \overline
D(\mathcal{JF}), \ \ \ \ \ \ \partial_2 {
\overline{\mathcal{F}}}=- \overline{\mathcal{F}}''-2 \overline D
D(\mathcal{J}\overline{\mathcal{F}}), \ea
which is recognized as the second flow of the a=4, N=4
supersymmetric KdV hierarchy \cite{dik}. Let us note that the N=2
superfield form of the seconf flow \p{n4-kdv} is related to its
component form \p{a4-kdv} as follows:
\ba \label{comp}
 &&u=1/2 ([D,\overline D]\mathcal{J}-\mathcal{J}')\Bigl
|,\ \ \ v=\mathcal{J}\Bigl |, \ \ \
 \xi=\overline D\mathcal{J}\Bigl |, \ \ \ \bar\xi= D\mathcal{J}\biggl |, \nn\\
&& r=\mathcal{F}\Bigl |, \ \ \ s= \overline{\mathcal{F}} \Bigl |,
\ \ \ \bar\eta= D  \overline{\mathcal{F}}\Bigl |, \ \ \
\eta=\overline D \mathcal{F}\Bigl |, \ea
where $|$ means the $(\theta,\bar \theta)\rightarrow 0$ limit. For
completeness, we give here the N=2 superfield form of the second
flow of the a=-2, N=4 super-KdV equation \p{a2-kdv}
\ba \label{n4-kdv2} \partial_2 \tilde {\mathcal{J}}&=&(
\tilde{\overline \mathcal{F}'}-\tilde\mathcal{F}'+
\tilde\mathcal{J}(\tilde\mathcal{F}+\tilde{\overline{\mathcal{F}}}))'
-2\overline D D (\tilde\mathcal{J}\tilde\mathcal{F})-2D\overline D  (\tilde\mathcal{J}\tilde{\overline{\mathcal{F}}}),\nn\\
\partial_2 \tilde\mathcal{F}&=&D\overline D  (\tilde\mathcal{J}'+1/2
\tilde\mathcal{J}^2 -\tilde\mathcal{F} \tilde{\overline \mathcal{F}}-3/2 \tilde\mathcal{F}^2),\nn\\
\ \partial_2 \tilde{\overline\mathcal{F}}&=&\overline D
D(-\tilde\mathcal{J}'+1/2 \tilde\mathcal{J}^2 -\tilde\mathcal{F}
\tilde{\overline \mathcal{F}}-3/2 \tilde{
\overline{\mathcal{F}}}{}^2),
  \ea
where the component content of the superfields is defined in the
same way as in eq.\p{comp}.

The precise analysis shows that
in addition to relation \p{1b} these exists at least one more
relation
\ba \label{2b}
 &&\mathcal{J}=-\frac{1}{2}\sum_{k=1}^2 F_k\overline F_k +\frac{
\overline{D}  F_2' }{\overline{D} F_2}+\left(
\frac{F_2\overline{D} F_1}{F_1\overline{D} F_2}\right )',\ \ \ \ \
\overline{\mathcal{F}}=-\overline{D}\left(\frac{F_2}{F_1}\right),\nn\\
&&
\mathcal{F}=D\left(\frac{1}{\overline{D}F_2}\left(\overline{D}\left(-F_1'+\frac{1}{4}F_1^2\overline
F_1+\frac{F_1'F_2\overline{D}F_1}{F_1\overline{D}F_2}
\right)+\frac{1}{2}F_2\overline F_2\overline{D}F_1\right)\right)
\ea
which connects the $(1,1)$-GNLS and a=4, N=4 super-KdV
hierarchies. Supplying  the supefields $F_a, \overline F_a$ in
eqs. \p{1b} and \p{2b} with the lattice index $i$ and $i-1$,
respectively, and equating the corresponding superfields $
\mathcal{J}, \mathcal{F}, \overline\mathcal{F}$ beloning to
relations \p{1b} and \p{2b}, we obtain the mapping
\ba \label{mapping}
 &&\frac{1}{2}\sum_{k=1}^2 (F_{k,i-1}\overline F_{k,i-1}-F_{k,i}\overline F_{k,i} )=
 \left(\log (\overline{D} F_{2,i-1}D\overline F_{2,i})+
\frac{F_{2,i-1}\overline{D} F_{1,i-1}}{F_{1,i-1}\overline{D}
F_{2,i-1}}\right )',\nn\\
&& \overline{D}D\left( \frac{\overline F_{1,i}}{D\overline
F_{2,i}}\right)=
\overline{D}\left(\frac{F_{2,i-1}}{F_{1,i-1}}\right)
\ ,\nn\\
&&
D\left(\frac{1}{\overline{D}F_{2,i-1}}\left(\overline{D}\left(F_{1,i-1}'-\frac{1}{4}F_{1,i-1}^2\overline
F_{1,i-1}-\frac{F_{1,i-1}'F_{2,i-1}\overline{D}F_{1,i-1}}{F_{1,-1}\overline{D}F_{2,i-1}}
\right)\right.\right.\nn\\
&&~~~~~~~~~~~~~~~~~~~~~~~~~~~~~~~~~~~~~~~~~~~~\left.\left.-\frac{1}{2}F_{2,i-1}\overline
F_{2,i-1}\overline{D}F_{1,i-1}\right)\right)=\frac12F_{1,i}D\overline
F_{2,i}\ , \ea
that acts like the discrete symmetry transformation of the
  (1,1)-GNLS hierarchy and has  significant
importance in our consideration. Namely, one can demonstrate that
all the  flows of the generalized N=4 supersymmetric TL hierarchy
\p{Lax} can be recovered from the corresponding flows of the
(1,1)-GNLS hierarchy with the help of the mapping \p{mapping}.


 Let us introduce the components of the N=2 superfields
entering into eq.\p{mapping} as follows:
\ba \label{cmp} && g_i=F_{1,i} \Bigl |, \ \  \bar g_i=\overline
F_{1,i} \biggl |, \ \ f_i=\overline D F_{2,i} \Bigl |, \ \
 \bar f_i= D\overline F_{2,i} \biggl |, \nn\\
&&\chi_i=F_{2,i} \Bigl |, \ \  \bar \chi_i=\overline F_{2,i}
\biggl |,\ \ \zeta_i=\overline D F_{1,i} \Bigl |, \ \  \bar
\zeta_i= D\overline F_{1,i} \biggl |,
 \ea
where the fields $g_i,\bar g_i, f_i,\bar f_i$  $(\chi_i, \bar
\chi_i,\zeta,\bar\zeta_i )$ are bosonic (fermionic) ones, and
define the following relations connecting these component fields
with the fields of the generalized N=4 supersymmetric TL hierarchy
in   basis \p{ab-basis}:
\ba\label{ab-gnls}
 &&a_i=-\frac{1}{2} (g_i\bar
g_i+\chi_i\bar\chi_i)+\left(\log f_i- \frac{\zeta_i\chi_i}{g_i
f_i}\right)',  \  \ \
 b_i=\frac{1}{2} (\frac{\bar
f_i\zeta_i\chi_i}{g_i}- f_i\bar f_i),\nn\\
&&\bar a_i=-\frac{1}{2} (g_i\bar g_i+\chi_i\bar\chi_i)+(\log
g_i)', \ \ \
 \bar b_i=\frac{1}{2} (\frac{
g_i\bar\zeta_i\bar\chi_i'}{\bar f_i}+ g_i\bar g_i'),\nn\\
&&\beta_i=\frac{1}{2}(f_i\bar\chi_i+\bar g_i \zeta_i)-\left(
\frac{\zeta_i}{g_i}\right )', \ \ \ \bar \beta_i=\frac{1}{2}(\bar
f_i
\chi_i- g_i \bar\zeta_i),\nn\\
&&\alpha_i=\frac{\bar \chi_i' }{\bar f_i}-\frac{\zeta_i}{g_i}, \ \
\ \bar \alpha_i =\frac{1}{f_i}(\chi_i'-\chi_i (\log
g_i)'+\frac{\zeta_i\chi_i\chi_i'}{g_if_i}).
 \ea

Now it is a matter of  straightforward calculations to verify that
the component form of the mapping \p{mapping} being rewritten in
  basis \p{ab-gnls} is equivalent to   the generalized N=4
supersymmetric TL equation \p{ab-toda}. Therefore, eq.\p{mapping}
represents the N=2 superfild form of this equation. Moreover, one
can verify that fields $\{a_i,\bar a_i, b_i,\bar b_i,
\chi_i,\bar\chi_i,\zeta_i,\bar\zeta_i \}$ defined by
eq.\p{ab-gnls} satisfy, as a consequence of eqs.\p{gnls-2f} and
\p{cmp}, the equations of the second flow \p{ba-2f} of the N=4
super-KdV hierarchy. Eliminating further the $x$-derivatives from
eq.\p{ba-2f} with the help of eq.\p{ab-toda} one obtains the
second flow \p{toda-2f} of the generalized N=4 supersymmetric TL
hierarchy in the basis \p{ab-basis}. It is clear that the same
procedure allows one to reproduce any flow of the generalized N=4
supersymmetric TL hierarchy \p{Lax} from the corresponding flow of
the (1,1)-GNLS hierarchy. Thus, we can reproduce the generalized
N=4 supersymmetric TL hierarchy starting with (1,1)-GNLS hierarchy
and mapping \p{mapping}.

Note that in a similar approach   the N=2 supersymmetric TL
hierarchy was constructed in \cite{BS}.
The N=2 supersymmetric TL hierarchy can be recovered from the
generalized N=4 supersymmetric TL hierarchy as a result of the
reduction with the reduction constraints \p{red_con}. The set of
equations \p{n2-toda} is exactly the first flow of this N=2
supersymmetric TL hierarchy. The N=4 supersymmetric differential
hierarchy in the basis \p{ab-basis}  can also be reduced to the
N=2 supersymmetric one by imposing the following reduction
constraints:
\ba \bar b=0,\ \ \ \bar a =-\frac{\beta \bar\beta}{b},\ \ \ \alpha
=\frac 1b (a\beta-\beta'),\ \ \ \bar\alpha= \frac
1b(\bar\beta(\log b)'-a\bar\beta-\bar\beta'). \ea
In this case, its second flow \p{ba-2f} takes the following
form(see eq.(50) in \cite{BS})
\ba
&&\partial_2 b=(-b''+2ba)'+2b(\frac{\beta\bar\beta}{b})',\ \ \ \
\partial_2\bar\beta=(\bar\beta'+2(a\bar\beta-\bar\beta(\log
b)'))',\nn\\
&&\partial_2
a=(a'+a^2+2(b+\beta(\frac{\bar\beta}{b})'+a\frac{\beta\bar\beta}{b}))',\
\ \ \
\partial_2 \beta=(-\beta'+2 a\beta)',\ \ \
 \ea
which is related to the second flow of the a=4, N=2 super-KdV
hierarchy \cite{Mat}.

The discrete symmetry \p{mapping} can equivalently be rewritten in
  terms of the N=2 superfields entering into the N=4 super-KdV
hierarchy. It reads\footnote{In \cite{sor4},  the N=4 superfield
form of the Darboux-Backlund symmetries of the N=4 super-KdV-Toda
hierarchy, which are related to eqs. \p{mapping} and \p{map2}, was
presented. }
\ba \label{map2}
&& \mathcal{F}_{i+1}
\overline{\mathcal{F}}_{i+1}=\Phi_i+D \overline{
D}\mathcal{J}_i-(\log \mathcal{F}_i)''+\frac{1}{\Phi_i}(\overline{
D}\mathcal{J}_i+\left(\frac{\overline{ D}
\mathcal{F}_i}{\mathcal{F}_i}\right)')(D\mathcal{J}_i'-D\mathcal{J}_i(\log
\mathcal{F}_i)'-\mathcal{F}_iD\overline{\mathcal{F}_i}),\nn\\
&&\mathcal{F}_{i}\overline{\mathcal{F}}_{i+1}=\Phi_i,\ \ \ \ \ \
\mathcal{J}_{i+1}=\mathcal{J}_i+(\log \Phi_i)', \ \ \ \ \ \ \ \ \
\Phi_i\equiv  \overline{ D}D\mathcal{J}_i+ \mathcal{F}_{i}
\overline{\mathcal{F}}_{i}+\frac{D\mathcal{J}_i\overline{ D}
\mathcal{F}_i}{\mathcal{F}_i}.
 \ea
 The discrete symmetries \p{mapping} and \p{map2} are useful when
constructing the solutions of the N=4 supersymmetric KdV and
(1,1)-GNLS  hierarchies. For example, if the set $\{
\mathcal{J}_i,\mathcal{F}_i,\overline\mathcal{F}_i\}$ is a
solution of the N=4 super-KdV hierarchy, then the set $\{
\mathcal{J}_{i+1},\mathcal{F}_{i+1},\overline\mathcal{F}_{i+1}\}$
is a solution of this hierarchy as well. Let us note that eqs.
\p{mapping} and \p{map2} set aside  boundary conditions.
Therefore, via eqs. \p{tr-a4} and \p{comp} one can obtain a
solution of the N=4 super-KdV hierarchy in terms of solutions of
the generalized N=4 supersymmetric TL hierarchy for different
boundary conditions including periodic ones \cite{my1,my2}.

\section{Conclusion}
In this paper, we  demonstrated that the generalized N=4
supersymmetric TL hierarchy contains the  N=4 supersymmetric KdV
one. We   used the first flow of the generalized N=4
supersymmetric TL hierarchy in order to express  all the lattice
fields in terms of the fields defined in the same lattice node and
rewrote   all its flows in the form of differential equations. In
such a way, we   reproduced the component form of the N=4 KdV
hierarchy as well as its supersymmetric transformations and
conservation laws. Finally, we   obtained two different N=2
superfield forms of the generalized N=4 supersymmetric TL equation
which are helpful when solving the N=4 super-KdV and (1,1)-GNLS
hierarchies.

In conclusion, we would like to note that the one-dimensional
generalized N=4 supersymmetric TL hierarchy is a particular case
of a wide class of hierarchies \cite{my3} defined by the Lax
operators
\ba\label{lll}
 (L)_{i,j}= \sum_{k=-2}^{2n}u_{k,i}
\delta_{i,j+k},\ \ \ \ u_{-2,i}=1,\ \ \ \ n>0. \ea
These hierarchies possess the N=2 supersymmetry and it would be
interesting to investigate which N=2 supersymmetric differential
hierarchies  can be reproduced from them in our approach.

Another problem  of special interest is to consider the Lax
operator \p{L-mat} with bosonic and fermionic fields replaced by
square $k\times k$ matrices with bosonic and fermionic entries,
respestively.   In our approach,  such a Lax operator gives rise
to the matrix hierarchy  which is N=4 supersymmetric and under
reduction constraints, when all the off-diagonal fields are equal
to zero, splits into $k$ independent N=4 super-KdV hierarchies.
One can expect that such matrix hierarchy can throw light on the
longstanding problem of constructing    the N=4
  super-KdV hierarchy (if any) with the N=4 $O(4)$ ("large")
superconformal algebra as the second Hamiltonian structure.

 \vspace{.8cm}
 {\bf Acknowledgments.  } The author is grateful to A.Sorin for
 clarifying remarks.

\end{document}